\documentclass[twocolumn,preprintnumbers,amsmath,amssymb]{revtex4}

\usepackage{epsfig}
\usepackage{subfigure}
\usepackage{graphics}
\usepackage{amsmath}
\usepackage{bbm}
\usepackage{hyperref}

\begin{document}

\title{Electric and magnetic response of hot QCD matter}
\author{T.~Steinert}%
\email{Thorsten.Steinert@theo.physik.uni-giessen.de}
\affiliation{%
 Institut f{\"u}r Theoretische Physik, %
  Universit\"at Giessen, %
  35392 Giessen, %
  Germany %
}

\author{W.~Cassing}
\affiliation{%
  Institut f{\"u}r Theoretische Physik, %
  Universit\"at Giessen, %
  35392 Giessen, %
  Germany %
}
\date{\today}
\begin{abstract}
We study the electric conductivity as well as the magnetic response
of hot QCD matter at various temperatures $T$ and chemical
potentials $\mu_q$ within the off-shell Parton-Hadron-String
Dynamics (PHSD) transport approach for interacting partonic systems
in a finite box with periodic boundary conditions. The response of
the strongly-interacting system in equilibrium to an external
electric field defines the electric conductivity $\sigma_0$ whereas
the response to a moderate external magnetic field defines the
induced diamagnetic moment $\mu_L$ ($T, \mu_q$) as well as the spin
susceptibility $\chi_S(T, \mu_q)$. We find a sizeable temperature
dependence of the dimensionless ratio $\sigma_0/T$ well in line with
calculations in a relaxation time approach for $T_c \! < \! T < \!
2.5 \!\, T_c$ as well as an increase of $\sigma_0$ with
$\mu_q^2/T^2$. Furthermore, the frequency dependence of the electric
conductivity $\sigma(\Omega)$ shows a simple functional form well in
line with results from the Dynamical QuasiParticle Model (DQPM). The
spin susceptibility $\chi_S(T,\mu_q)$ is found to increase with
temperature $T$ and to rise $\sim \mu_q ^2/T^2$, too. The actual
values for the magnetic response of the QGP in the temperature range
below 250 MeV show that the QGP should respond diamagnetically in
actual ultra-relativistic heavy-ion collisions since the maximal
magnetic fields created in these collisions are smaller than
$B_c(T)$ which defines a boundary between diamagnetism
and paramagnetism. \\

%
%
\end{abstract}
\maketitle
\section{Introduction}
The phase diagram of strongly interacting hadronic/partonic matter
has been a subject of primary interest in the physics community for
decades. At vanishing (or low) chemical potentials lattice QCD
(lQCD) calculations have provided reliable results  on the equation
of state \cite{lQCD,Peter} and given a glance at the transport
properties in particular in the partonic phase. On the other hand
high energy heavy-ion reactions are studied experimentally and
theoretically to obtain information about the properties of nuclear
matter under the extreme conditions of high baryon density and/or
temperature. Ultra-relativistic heavy-ion collisions at the
Relativistic Heavy-Ion Collider (RHIC) and the Large Hadron Collider
(LHC) at CERN have produced a new state of matter, the strongly
interacting quark-gluon plasma (sQGP), for a couple of fm/c in
volumes up to a few $10 ^3$ fm$^3$ in central reactions. The
produced QGP shows features of a strongly-interacting fluid unlike a
weakly-interacting parton gas \cite{StrCoupled1} as had been
expected from perturbative QCD (pQCD). Large values of the observed
azimuthal asymmetry of charged particles in momentum
space~\cite{STAR,PHENIX,BRAHMS,PHOBOS,ALICE}, i.e. the elliptic flow
$v_2$ could quantitatively be well described by ideal hydrodynamics
up to transverse momenta of the order of 1.5~GeV/c
\cite{IdealHydro1,IdealHydro2,IdealHydro3,IdealHydro4,IdealHydro5,IdealHydro6}.
Recent studies of 'QCD matter' in equilibrium -- using lattice QCD
calculations \cite{l0,ll0} or partonic transport models in a finite
box with periodic boundary conditions \cite{Vitalii1,Vitalii2} --
have demonstrated that the ratio of the shear viscosity to entropy
density $\eta/s$ should have a minimum close to the critical
temperature $T_c$, similar to atomic and molecular systems
\cite{review}. On the other hand, the ratio of the bulk viscosity to
the entropy density $\zeta/s$ should have a maximum close to $T_c$
\cite{Vitalii2} or might even diverge at $T_c$
\cite{MaxBulk1,MaxBulk2,MaxBulk3,MaxBulk4,MaxBulk5}. Indeed, the
minimum of $\eta/s$ at $T_c \approx$ 160 MeV is close to the lower
bound of a perfect fluid with $\eta/s= 1/(4\pi)$~\cite{KSS} for
infinitely coupled supersymmetric Yang-Mills gauge theory (based on
the AdS/CFT duality conjecture). This suggests the `hot QCD matter'
to be the `most perfect fluid'~\cite{new1,new2,Barbara}. On the
empirical side, relativistic viscous hydrodynamic calculations
(using the Israel-Stewart framework) also require a very small
$\eta/s$ of $0.08-0.24$ in order to reproduce the RHIC elliptic flow
$v_2$ data
\cite{ViscousHydro1,ViscousHydro2,ViscousHydro3,ViscousHydro4};
these phenomenological findings thus are in accord with the
theoretical studies for $\eta/s$ in
Refs.~\cite{Vitalii2,Mattiello,Greco}.

Whereas shear and bulk viscosities of hot QCD matter at finite
temperature $T$ presently are roughly known, the electric
conductivity $\sigma_0$ is a further macroscopic quantity of
interest~\cite{Hirono:2012rt,Jorge} since it controls the
electromagnetic emissivity of the plasma.  First results from
lattice calculations on the electromagnetic correlator have provided
results that varied by more than an order of magnitude
\cite{l1,l2,l3,l4,l5}. Furthermore, the conductivity dependence on
the temperature $T$ (at $T\!\!>\!\!T_c$) is widely unknown, too. The
electric conductivity $\sigma_0$ is also important for the creation
of electromagnetic fields in ultra-relativistic nucleus-nucleus
collisions from partonic degrees-of-freedom, since $\sigma_0$
specifies the imaginary part of the electromagnetic (retarded)
propagator and leads to an exponential decay of the propagator in
time $\sim \! \exp(-\sigma_0 (t-t')/({\hbar} c))$ \cite{Tuchin}.

Apart from the electric conductivity the magnetic response of the
QGP (or strong vacuum) to  external magnetic fields ${\bf B}$ has also been of current
interest from the experimental side
\cite{CME1,CME2,CME3,CME4,CME5,CME6} as well as from lattice
QCD \cite{lqcdb0,lqcdb1,lqcdb2,lqcdb3}. Strong magnetic fields are
created in peripheral relativistic nucleus-nucleus collisions by
the charges of the spectator protons during the passage time of
the nuclei \cite{CME6,CME7,Slava1} and at the top RHIC energy
of $\sqrt{s_{NN}}$ = 200 GeV magnetic fields of order $eB \approx
5 m_\pi^2$ can be reached. This has
lead to the suggestion of a charge separation effect due to the Chiral-Magnetic-Effect (CME) in
these reactions \cite{CME1,CME2,CME3,CME4,CME5,CME6}. On the other
side lQCD has been focusing on the magnetic catalysis of
the chiral $<q{\bar q}>$ condensate at very strong ${\bf B}$ fields at low
temperature and the inverse  magnetic catalysis of
the chiral condensate at temperatures close the critical
temperature indicating a decrease of $T_c$ for high magnetic
fields. Note, however, that these studies involve time-independent
magnetic fields that are significantly higher than those achieved
in peripheral nucleus-nucleus collisions for very short times where an
equilibrated QGP might not have been established.
Nevertheless, a sufficient knowledge of the electric and magnetic response
of the QGP (in equilibrium) to external
electromagnetic fields  is mandatory to
explore a possible generation of the Chiral-Magnetic-Effect  in
predominantly peripheral heavy-ion reactions
\cite{CME1,CME2,CME3,CME4,CME5,CME6} and to determine the photon production
from the QGP in heavy-ion collisions at different
centralities and bombarding energies \cite{photon1,photon2,photon3}.

In this work we extend our previous studies on the electric
conductivity $\sigma_0(T)$ \cite{Ca13,Marty} for `infinite parton matter'
also to finite quark chemical potential $\mu_q$ employing the
Parton-Hadron-String Dynamics (PHSD) transport approach
\cite{PHSD1}, which is based on generalized transport equations
derived from the off-shell Kadanoff-Baym equations \cite{Kadanoff1,Kadanoff2}
for Green's functions in phase-space representation (beyond the
quasiparticle approximation). This approach describes the full
evolution of a relativistic heavy-ion collision from the initial
hard scatterings and string formation through the dynamical
deconfinement phase transition to the strongly-interacting
quark-gluon plasma (sQGP) as well as hadronization and the
subsequent interactions in the expanding hadronic phase. In the
hadronic sector PHSD is equivalent to the Hadron-String-Dynamics
(HSD) transport approach \cite{CBRep98,Brat97,Cass02,PRL03} -- a covariant extension of
the Boltzmann-Uehling-Uhlenbeck (BUU) approach~\cite{Cass90} -- that
has been used for the description of $pA$ and $AA$ collisions from
lower Schwerionen-Synchrotron (SIS) to RHIC energies in the past. On the other hand, the
partonic dynamics in PHSD is based on the Dynamical Quasi-Particle
Model (DQPM) \cite{DQPM1,DQPM2,DQPM3}, which describes QCD
properties in terms of single-particle Green's functions (in the
sense of a two-particle irreducible (2 PI) approach) and reproduces
lattice QCD results -- including the partonic equation of state --
in thermodynamic equilibrium. For further details on the PHSD
off-shell transport approach and hadronization we refer the reader
to Refs.~\cite{PHSD1,Cassing,Bratkovskaya:2011wp,Vitalii1}.

The layout of our study is as follows: In Section II we concentrate
on calculating the electric conductivity  for `infinite' QCD matter
also at finite quark chemical potential $\mu_q$  and provide simple
parametrizations for the dependence of $\sigma_0$ on $\mu_q$.
Furthermore, we calculate the frequency dependence $\sigma(\Omega)$
for periodic external fields and compare the results with those from
the Dynamical QuasiParticle Model (DQPM). In Section III we compute
the diamagnetic and paramagnetic contributions to the magnetization
$M$ of the plasma as a function of temperature $T$ and $\mu_q$  and
compare to the experimental situation at RHIC. A summary and
discussion of results is presented in Section IV.

\section{Electric conductivity}
We briefly recall the setup of our studies within PHSD. The 'infinite' hadronic or QCD
matter is simulated within a cubic box with periodic boundary
conditions at various values for the energy density (or temperature)
and the quark chemical potential $\mu_q$. The size of the box is
fixed to $V\!=\!9^3$ fm$^3$ as in the previous investigations
\cite{Vitalii1,Vitalii2,Ca13}. The initialization is done by
populating the box with light ($u,d$) and strange ($s$) quarks,
antiquarks and gluons slightly out of equilibrium. The system
approaches kinetic and chemical equilibrium during its time
evolution within PHSD.  For more details on the simulation of
equilibrated partonic systems using PHSD in the box at finite temperature
$T$ and quark chemical potential $\mu_q$ we refer the
reader to Ref.~\cite{Vitalii1}.

We recall that PHSD is an off-shell transport approach that
propagates quasi-particles with broad spectral functions.
Numerically, the continuous spectral distribution in the mass of a
particle (given by its spectral function) is probed by a large
number of test-particles with (evolving) masses $M_j(t)$. In order
to include the effects from an external electric field ${\bf E}$ or
magnetic field ${\bf B}$, the propagation of each charged
test-particle $j$ is performed with the additional Lorentz force in
the equation of motion:
\begin{equation} \label{e1}
\frac{d}{dt} {\bf p}^j = q_j e ({\bf E} + \frac{{\bf p}^j}{E^j} \times {\bf B}),
\end{equation}
where $q_j$ denotes the fractional charge of the test-particle
($\pm 1/3, \pm 2/3$) and $E^j$ its energy. We recall that the
external electric field will lead to an acceleration of positively
and negatively charged particles in opposite directions while the
particle scatterings/interactions will damp this acceleration and
eventually lead to an equilibrium current (cf. Fig. 1 in Ref.
\cite{Ca13}).
The  electric current density $j_z(t)$ (for an external
electric field in $z$-direction) is calculated by
\begin{equation} \label{e2} j_z(t) = \frac{1}{VN} \sum_{k=1}^N \sum_{j=1}^{N_k(t)} \ e q_j
\frac{p_z^j(t)}{M_j(t)}, \end{equation}
 where $M_j(t)$ is the mass
of the test-particle $j$ at time $t$. The summation in (\ref{e2}) is
carried out over $N$ ensemble members $k=1 \dots N$ while $N_k(t)$
denotes the time-dependent number of 'physical' ($u,d,s$) quarks and
antiquarks that varies  with time $t$ due to the processes $q +
\bar{q} \leftrightarrow g \leftrightarrow q'+{\bar q}'$ in a single
member of the ensemble (run). The number of runs $N$ is typically
taken as a few hundred which gives a current $j_z(t)$ practically
independent on the number of ensemble members $N$. We recall that
(without external fields) each run of the ensemble is a
microcanonical simulation of the dynamics as inherent in the PHSD
transport approach which strictly conserves the total four-momentum
as well as all discrete conservation laws (e.g. net fermion number
for each flavor etc.) \cite{Vitalii1}. A note of caution has to be
given, since due to an external field we deal with an open system
with increasing energy density (temperature) in time. Therefore we
employ sufficiently small external fields $eE_z$, such that the
energy increase during the computation time (in each run) stays
below 2\% and the increase in temperature below 1 MeV.

\subsection{Constant electric fields}
\begin{figure}[hbt]
   \centering
   \includegraphics[width=0.95\linewidth]{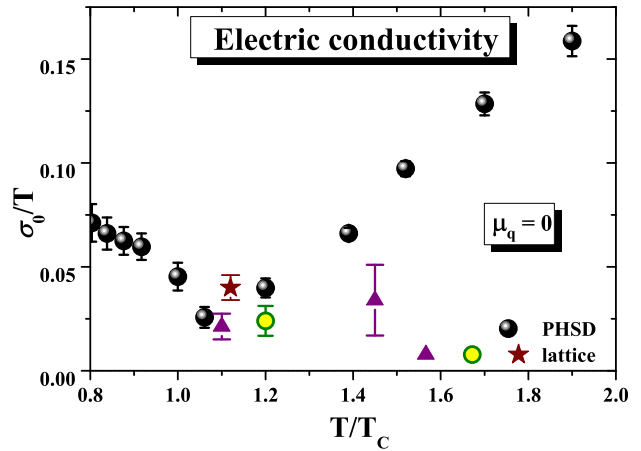}
   \caption{(Color online) The dimensionless ratio of electric conductivity over temperature
   $\sigma_0/T$ (\ref{e4})    as a function of the scaled temperature
   $T/T_c$ for $\mu_q=0$ in comparison to recent lattice QCD results: triangles -
   quenched QCD results in the continuum
   limit with Wilson-Clover fermions and renormalized vector currents \cite{l1}, star -
   quenched SU(2) lattice gauge
   theory \cite{l4}, open circle - QCD with two dynamical flavors of Wilson-Clover fermions \cite{l5}. The PHSD
   results (full dots) are the same as in Ref. \cite{Ca13}. }
   \label{figg1}
\end{figure}

We find that for constant electric fields up to $e E_z$ = 50
$\text{MeV}/\text{fm}$ a stable electric current $j_{eq}$ emerges
that is $\sim E_z$ (cf. Ref. \cite{Ca13}). Accordingly,  we obtain
the conductivity $\sigma_0(T,\mu_q)$ from the ratio of the
stationary current density $j_{eq}$ and the electric field strength
as
\begin{equation} \label{e4}
\frac{\sigma_0(T,\mu_q)}{T} = \frac{j_{eq}(T,\mu_q)}{E_z T} \ .
\end{equation}
The results for the dimensionless ratio (\ref{e4}) at $\mu_q=0$
(from Ref. \cite{Ca13}) are displayed in Fig. \ref{figg1} as a
function of the scaled temperature $T/T_c$ in comparison to more
recent lattice QCD results from Refs. \cite{l1,l4,l5} and suggest a
minimum in the ratio $\sigma_0(T,\mu_q=0)/T$ close to the critical
temperature $T_c$ followed by an approximate linear rise up to 2
$T_c$ (cf. Ref. \cite{Ca13}). The most recent lQCD results
\cite{l1,l4,l5} are roughly compatible with the PHSD calculations.
Moreover, the lattice data from Ref. \cite{Aarts} give support for
the linearity of $\sigma_0/T$ with temperature above $T_c$.

\begin{figure}[htb]
  \centering
  \includegraphics[width=0.95\linewidth]{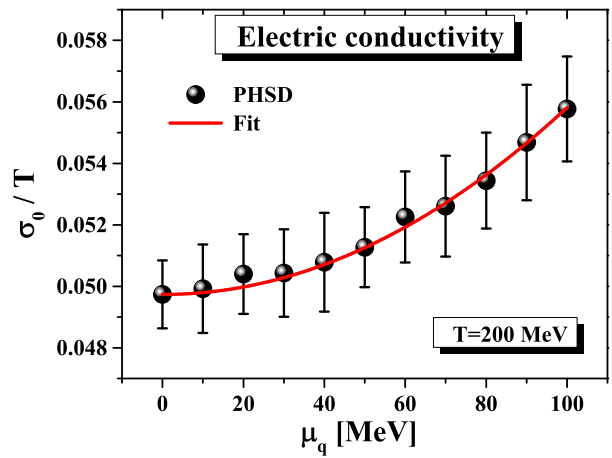}
  \caption{(Color online) The electric conductivity over temperature $\sigma_0 / T$ as a
  function of the chemical potential $\mu_q$ for $T$= 200 MeV
  from the PHSD calculations (full dots).
  The error bars indicate the statistical uncertainty for the ratio (\ref{e4}) when performing calculations
  for different external field strength $eE_z$ up to 50 MeV/fm.}
  \label{pic:condTvsmu}
\end{figure}

We now focus on the explicit dependence of $\sigma_0(T,\mu_q)/T$ as a
function of the chemical potential $\mu_q$ which is shown in
Fig.~\ref{pic:condTvsmu}  for a fixed temperature $T$=200 MeV.
The numerical result can be fitted with a quadratic correction (solid line in
Fig.~\ref{pic:condTvsmu})
\begin{equation} \label{parabol}
\frac{\sigma_0(T,\mu_q)}{T} =  \frac{\sigma_0(T,\mu_q=0)}{T} \left(1
+ a(T) \mu_q^2 \right) .
\end{equation} with $a(T) \approx 11.6 \ \text{GeV}^{-2}$ for $T=0.2$ GeV.
This result comes about as follows: We recall that the electric
conductivity of gases, liquids and solid states is described in the
relaxation time approach by the Drude formula
\begin{equation} \label{eq7} \sigma_0 = \frac{e^2 n_e \tau}{m_e^*} ,
\end{equation}
where $n_e$ denotes the density of non-localized charges, $\tau$ is
the relaxation time of the charge carriers in the medium and $m_e^*$
their effective mass. This expression can be directly computed for
partonic degrees-of-freedom within the DQPM, which was used to match
in PHSD the quasiparticles properties to lattice QCD results in
equilibrium for the equation-of-state (EoS) as well as various
correlators \cite{DQPM1,DQPM2,DQPM3}. We note that the
electromagnetic correlator from lQCD calculations \cite{l1} appears
to match rather well the back-to-back dilepton rate from PHSD at
$T=1.45 T_c$ (cf. Fig.~2 in Ref.~\cite{Olena13}), which suggests that the
results of our calculations for $\sigma_0$ - for vanishing invariant mass -
should also be close to the lQCD extrapolations from~\cite{l1}.

In the DQPM, the relaxation time for quarks/antiquarks is given by
$\tau = 1/\Gamma_q(T,\mu_q)$, where $\Gamma_q(T,\mu_q)$ is the width
of the quasiparticle spectral function
(cf.~\cite{DQPM1,Bratkovskaya:2011wp}). Furthermore, the spectral
distribution for the mass of the quasiparticle has a finite pole
mass $M_q(T,\mu_q)$ that is also fixed in the DQPM, as well as the
density of ($u, \bar{u}, d, \bar{d}, s, \bar{s}$) quarks/antiquarks
as a function of temperature and chemical potential (cf.
Refs.~\cite{DQPM1,Bratkovskaya:2011wp}). Thus, we obtain for the
dimensionless ratio (\ref{e4}) the expression~\cite{Ca13}
\begin{equation} \label{e8}
\frac{\sigma_0 (T,\mu_q)}{T} \approx \frac{2}{9} \frac{e^2
n_{q+{\bar q}}(T,\mu_q)}{M_q(T,\mu_q) \Gamma_q(T,\mu_q) T} ,
\end{equation}
where $n_{q+{\bar q}}(T,\mu_q)$ denotes the total density of quarks and
antiquarks and the prefactor $2/9$ reflects the flavor averaged
fractional quark charge squared $(\sum_f q_f^2)/3$. As found in our
previous study \cite{Ca13} the DQPM results match well with the
explicit PHSD calculations in the box for $\mu_q$=0 since  PHSD in
equilibrium is a suitable transport realization of the
DQPM~\cite{Vitalii1}.

\begin{figure}[htb]
  \centering
  \includegraphics[width=0.95\linewidth]{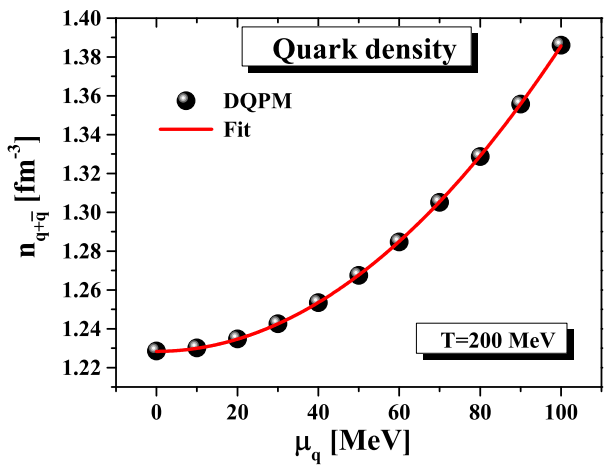}
  \caption{(Color online) Quark+antiquark density from the DQPM (full dots)
  as a function of the quark chemical potential $\mu_q$
  for $T=200$ MeV. The solid (red) line  displays the fit (\ref{D5}) to the DQPM results.}
  \label{pic:densityfinitmu}
\end{figure}

In the DQPM we have $\Gamma_q(T,\mu_q) \approx \Gamma_q(T,\mu_q =0)$
and $M_q(T,\mu_q) \approx M_q(T,\mu_q=0)$ for $\mu_q \leq $ 100 MeV,
however, \begin{equation} \label{D5} n_{q+{\bar q}}(T,\mu_q) \approx
n_{q+{\bar q}}(T,\mu_q=0) \left( 1 + a(T) \mu_q^2 \right)
\end{equation} with the same coefficient $a(T)$ as in Eq.
(\ref{parabol}). This is demonstrated explicitly in Fig.
\ref{pic:densityfinitmu} where the actual DQPM results for the
quark+antiquark density (full dots) are compared to the  fit
(\ref{D5}) (solid line).

The temperature dependence of the expansion coefficient $a(T)$ is
found to be $\sim 1/T^2$ such that  the ratio $\sigma_0/T$ can be approximated by
\begin{equation} \label{expand}
\frac{\sigma_0(T,\mu_q)}{T} \approx  \frac{\sigma_0(T,\mu_q=0)}{T}
\left(1 + c_{\sigma_0} \frac{\mu_q^2}{T^2} \right) .
\end{equation}

\begin{figure}[h!bt]
   \centering
   \includegraphics[width=0.95\linewidth]{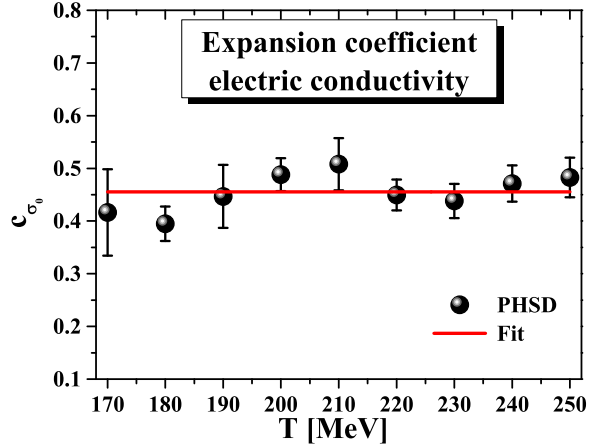}
   \caption{(Color online) The expansion factor $c_{\sigma_0}$ in (\ref{expand}) as a function of the temperature
   $T$ for $\mu_q \leq$ 100 MeV. The solid line shows the average value in the interval 170 MeV $ < T < $ 250 MeV.}
   \label{pic:korrfaktor}
\end{figure}

In Fig. \ref{pic:korrfaktor} we display the coefficient
$c_{\sigma_0}$ in the temperature range 170 MeV$ \leq T \leq $ 250
MeV giving $c_{\sigma_0} \approx 0.46$ as a best fit. This
completes our study on the stationary electric conductivity
$\sigma_0$.

\subsection{Periodic electric fields}
\begin{figure}[htb]
  \centering
  \includegraphics[width=0.95\linewidth]{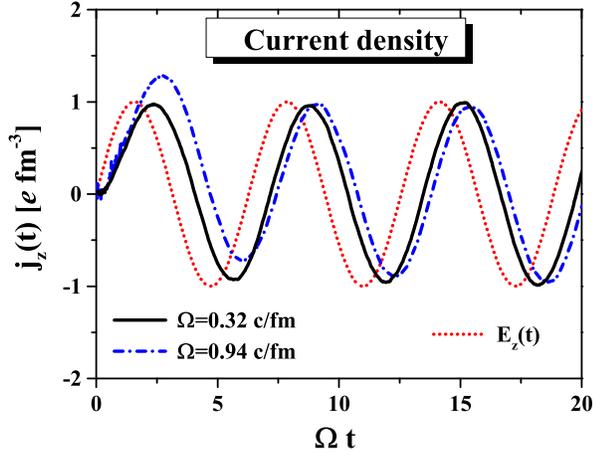}
  \caption{(Color online) The time-dependent electric current density $j_z(t)$ for $\Omega=0.32 \
  \text{c}/\text{fm}$ (solid black) and $\Omega= 0.94 \ \text{c}/\text{fm}$ (dash dotted blue) normalized to the
  equilibrium  amplitude for temperature $T=190$ MeV and $e E_z^0=0.005 \ \text{GeV}^2 \approx 25$ MeV/fm.
  The dotted red line shows the time-dependence
  of the external electric field $E_z(t)$.}
  \label{pic:jzvsw}
\end{figure}

\begin{figure}[htb]
  \centering
  \includegraphics[width=0.95\linewidth]{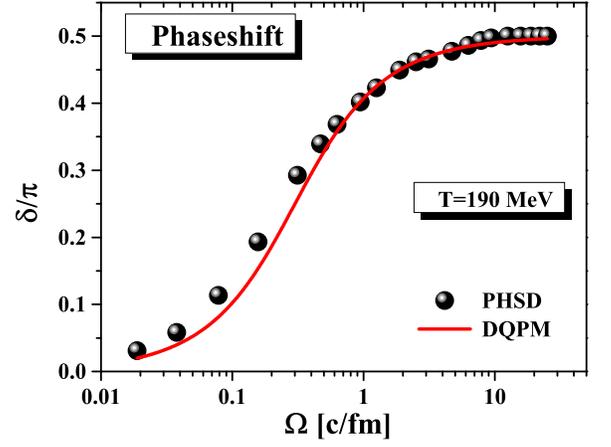}
  \caption{(Color online) The phase shift $\delta$ over $\pi$ as a function of the frequency $\Omega$
  from the PHSD calculations (full dots) for $T=190$ MeV.
  The red line shows the phase shift as expected from the DQPM using (\ref{sigo}).}
  \label{pic:phasevsw}
\end{figure}

\begin{figure}[htb]
  \centering
  \includegraphics[width=0.95\linewidth]{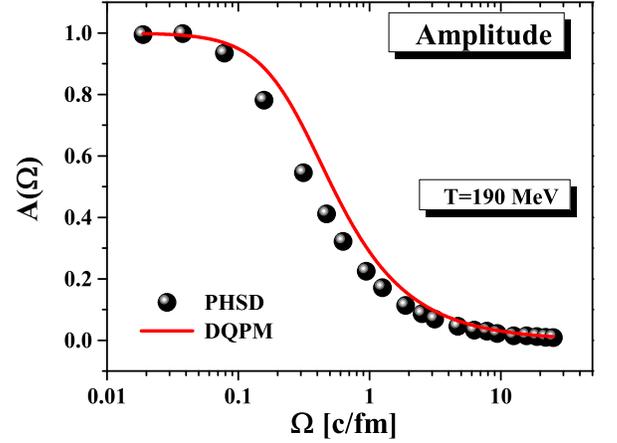}
  \caption{(Color online) The amplitude $A(\Omega)$   as a
  function of the frequency $\Omega$ from the PHSD calculations (full dots) for $T=190$ MeV. The red
  line shows the expected amplitude within the DQPM.}
  \label{pic:ampvsw}
\end{figure}

We now extent our study to
external periodic fields of frequency $\Omega$,
\begin{equation}
 E_z(t)=E_z^0 \sin (\Omega t).
\end{equation}
In this case the electric current density $j_z(t)$ does not achieve
a constant equilibrium value and also oscillates with the frequency
$\Omega$. Fig. \ref{pic:jzvsw} shows the time-dependence of the
current $j_z(t)$ from PHSD for different frequencies as a function
of $\Omega t$ with their amplitudes normalized to one in comparison
to the external electric field $E_z(t)$ (dotted red line).
The current $j_z(t)$ is seen to be shifted in phase compared to the
electric field; the phase shift $\delta$  increases with the
frequency $\Omega$ up to $\pi/2$. The currents in Fig.
\ref{pic:jzvsw} can be well described by
\begin{equation}
j_z(t)= A(\Omega) j_{eq} \sin(\Omega t-\delta(\Omega)).
\end{equation}
We find that the amplitude $A(\Omega)$ decreases with the frequency
$\Omega$ since the current has less time to build up and to follow
the external field. This behavior is in line with the complex
conductivity $\sigma(\Omega)$ for oscillating fields,
\begin{equation} \label{sigo}
\sigma(\Omega)=\frac{\sigma_0}{1- i {\Omega}/\Gamma_q} =
\frac{\sigma_0}{1+  {\Omega^2}/\Gamma_q^2} + i \frac{\sigma_0
\Omega/\Gamma_q}{1+ {\Omega^2}/\Gamma_q^2} ,
\end{equation}
where $\Gamma_q$ is the quasi-particle width of the charged
particles (quarks and antiquarks). We have computed the current
$j_z(t)$ for $T=190$ MeV and $e E_z^0=0.005 \ \text{GeV}^2 \approx
25$ MeV/fm in the frequency range $0.02 \ \text{c}/\text{fm} < \Omega
< 25 \ \text{c}/\text{fm}$. Fig. \ref{pic:phasevsw} shows the phase
shift $\delta = \arctan(\Omega/\Gamma_q)$ and Fig. \ref{pic:ampvsw}
the amplitude $A(\Omega) = 1/\sqrt{1+\Omega^2/\Gamma_q^2}$ (full
dots). The PHSD results can be easily followed up within the DQPM
results (shown by the red lines) which provide again a good
description of the microscopic calculations. Since the complex
conductivity $\sigma(\Omega)$ depends only on the width $\Gamma_q$
 and the stationary conductivity $\sigma_0$ in (\ref{sigo}) its actual
values for different temperatures $T$ and finite chemical potential
$\mu_q$ follow directly from our previous results in this Section.
Note that for actual electric fields in peripheral Au+Au collisions
at the top RHIC energy we have $\Omega \approx 22 \
\text{c}/\text{fm}$ such that the electric conductivity
$\sigma(\Omega)$ is suppressed relative to its equilibrium value by
more than a factor of 100.

\section{Magnetic response}
In order to explore the magnetic response of the partonic system
within PHSD we will assume the magnetic field to be sufficiently
small such that terms $\sim B^2$ can be neglected (see below). Note that this limit does
not hold for the strong fields $eB (\sim $ 0.1-1 GeV$^2 \approx$ 0.5-5 GeV/fm)
in actual lattice QCD studies \cite{lqcdb1,lqcdb2,lqcdb3}. Using
\begin{equation}
 (\boldsymbol{\sigma} \textbf{D})^2=\textbf{D}^2-q e \boldsymbol{\sigma}\cdot \textbf{B}, \
  \textbf{D}^2=(\textbf{p}-q
e \textbf{A})^2=\textbf{p}^2-q e \textbf{L} \cdot \textbf{B}
\end{equation} with the Pauli matrices $\boldsymbol{\sigma}$, the
kinetic momentum ${\bf p}$ and the angular momentum ${\bf L}$ the
Dirac equation can be rewritten for 2-component quark and antiquark
spinors leading to the Hamiltonian
\begin{equation}\label{eq:hamiltondirac}
 H_{Dirac}=\sqrt{\textbf{p}^2+m^2-q e (\textbf{L}+\boldsymbol{\sigma}) \cdot \textbf{B}} \end{equation} $$
 \approx E - \frac{qe}{2 E}
(\textbf{L}+\boldsymbol{\sigma}) \cdot \textbf{B}=E - \frac{qe}{2 E}
(\textbf{L}+2\textbf{S}) \cdot \textbf{B}  $$
with $E= \sqrt{{\bf
p}^2+m^2}$. In case of small energies
 $E \rightarrow
\frac{\textbf{p}^2}{2m}+m$  this leads to the well known expression
for the non-relativistic  Pauli equation:
\begin{equation}
 H_{Pauli}=\frac{\textbf{p}^2}{2 m} - \frac{qe}{2 m} (\textbf{L} + \boldsymbol{\sigma})
\cdot \textbf{B}.
\end{equation}
The change of the energy of the system in the presence of an
external magnetic field ${\bf B}$ is determined by the magnetic moment
$\boldsymbol{\mu}$:
\begin{equation}\label{eq:magmoment}
 \boldsymbol{\mu}=\boldsymbol{\mu}_L+\boldsymbol{\mu}_S=\frac{qe}{2 E} (\textbf{L} +
 2\textbf{S}) ,
\end{equation}
which has a contribution from the angular momentum ${\bf L}$ of a
particle and from the spin ${\bf S}= \boldsymbol{\sigma}/2$. In the
following we will investigate both terms separately since they
provide contributions to the magnetic moment of opposite sign.
In analogy to Sec. II we are dealing with an open system but the increase
in the total energy stays below 1\%.

\subsection{Diamagnetic contribution}
The induced angular momentum ${\bf L}$ emerges from the Lorentz
force (\ref{e1}) on a charged particle due to an external field
${\bf B}$,
\begin{equation}
 \textbf{F}_L=\frac{q e}{E} (\textbf{p} \times \textbf{B}),
\end{equation}
and induces a magnetic moment opposite to the direction of the ${\bf
B}$-field since the charged particle spirals around the magnetic
field with frequency $\omega=\frac{q e B}{E}= \frac{p_{\perp}}{E
R}$, where $p_{\perp}$ is the momentum component of the particle
perpendicular to the direction of the magnetic field and $R$ is the
radius of the spiral. We obtain alternatively for the angular
momentum
\begin{equation}
 \textbf{L}=\frac{R q e}{|\textbf{F}_L| E} (\textbf{p} (\textbf{p} \cdot \textbf{B})-\textbf{B}
 \textbf{p}^2) ,
\end{equation} where
$\textbf{p} (\textbf{p} \cdot \textbf{B})$ is the projection of the
momentum on the direction of the magnetic field $\textbf{e}_B$.
Inserting the expression for the radius $R$ we get
\begin{equation}
 \textbf{L}=\frac{-p_{\perp}^3}{|\textbf{F}_L| E} \text{sign}(q)
 \textbf{e}_B \ .
\end{equation}
Assuming the magnetic field to be oriented in $y$-direction and
employing the Lorentz force  $|\textbf{F}_L|=\frac{|q e B|}{E}
p_{\perp}$ we end up with
\begin{equation}\label{eq:drehimpuls}
 L_y=\frac{-p_{\perp}^2}{q e B}.
\end{equation}
This gives the induced magnetic moment
\begin{equation}\label{eq:magmomLklassisch}
 \mu_L=\frac{-p_{\perp}^2}{2 B E}.
\end{equation}
Since the Lorentz force changes only the direction of ${\bf p}$ and
not its magnitude $|{\bf p}|$ the particle energy $E$ is conserved,
too. As a consequence the energy contribution in the Hamiltonian
(\ref{eq:hamiltondirac}) is independent from the magnetic field
strength:
\begin{equation} \label{e9}
\Delta E_{mag,L}=-\mu_L B=-\frac{-p_{\perp}^2}{2 B E}
B=\frac{p_{\perp}^2}{2 E}.
\end{equation}
We note that the diamagnetic contribution can not be seen in
approaches that calculate the magnetization by differentiation of
the thermodynamic potential (e.g. free energy F) with respect to
the magnetic field $B$.
In principle, the angular momentum ${\bf L}$ has to be quantized.
However, the actual values for ${\bf L}$ (in units of $\hbar$) are
$\gg$ 1 for 'small' field strength since $|{\bf L}| \sim 1/(eB)$
such that quantum corrections are subleading in our case.

\subsection{Paramagnetic contribution}
The quark and antiquark spins provide a paramagnetic contribution
since the spin precession around the direction of the magnetic field
${\bf B}$ in thermal equilibrium gives a positive magnetic moment
$\mu_S$ since the energy becomes reduced according to Eq.
(\ref{eq:hamiltondirac}). The spin degree-of-freedom is
introduced in PHSD in line with the generalized test-particle ansatz
\cite{Kadanoff2} for the Wightman function
\begin{equation}\label{eq:testspin}
i G^<(X,P,S) = \frac{1}{N} \sum_{k=1}^N \sum_{i=1}^{N_k(t)}
\delta^{(3)}(\textbf{X}-\textbf{X}_i(t)) \end{equation} $$ \times
\delta^{(3)}(\textbf{P}-\textbf{P}_i(t))
\delta(P_0-\epsilon_i(t))\delta^{(2)}(\textbf{S}-\textbf{S}_i(t)) $$
where $X$ and $P$ stand for space-time and four-momentum
coordinates, respectively, while ${\bf S}$ denotes the spin
degree-of-freedom. In (\ref{eq:testspin}) the number of ensemble
members (runs) is denoted by $N$ whereas $N_k$ is the number of
partons in the run $k= 1 \dots N$ that describe the 'physical'
particles in each microcanonical simulation. The spin
degree-of-freedom has to be treated in line with quantum mechanics
according to the interaction Hamiltonian (\ref{eq:hamiltondirac}),
i.e.
\begin{equation} \label{HS}
 {\hat H}_S=- \frac{qe}{2 E} \boldsymbol{\sigma} \cdot  \textbf{B}.
\end{equation}
The spin-wavefunction for a spin $1/2$ fermion is taken as a
2-component spinor
\begin{equation}
 |\chi\rangle=\left|
 \begin{aligned}
  \uparrow \\ \downarrow
 \end{aligned}
 \right>
\end{equation}
with $\langle \uparrow|\uparrow \rangle$ denoting the probability for the spin in
$z$-direction (parallel to the magnetic field) while
$\langle \downarrow|\downarrow \rangle$ stands for the probability for the
anti-parallel orientation. We assume the spin-wavefunction to be
normalized. i.e.
$\langle \chi|\chi \rangle=\langle \uparrow|\uparrow \rangle+\langle \downarrow|\downarrow \rangle=1$. The
projection on the coordinate axis $i$ is provided by
\begin{equation} \label{eq:projection}
 S_i=\frac{1}{2} \langle \chi|\sigma_i|\chi \rangle.
\end{equation}
The time-evolution of the spin projections according to the
Hamiltonian (\ref{HS}) can be worked out in a straight forward way using
\begin{equation}
 |\chi(t)\rangle={\hat U}(t,t_0)|\chi(t_0) \rangle =e^{-i{\hat H}_S (t-t_0)}|\chi(t_0)\rangle
\end{equation} with
\begin{equation}
 {\hat U}(t,t_0)
 =\mathbbm{1}_2 \cos(\frac{qe}{2 E} B (t-t_0))+i\frac{\boldsymbol{\sigma} \cdot
 \textbf{B}}{B} \sin(\frac{qe}{2 E} B(t-t_0)).
\end{equation}
and Eq. (\ref{eq:projection}).
The resulting equations of motion lead to a precession of the spin
of a quark/antiquark with frequency $\omega = q e B/E$ which changes
only if the energy of the particle $E$ changes in a collision or in
the inelastic reaction $q + {\bar q} \rightarrow g \rightarrow q' +
{\bar q}'$.
Since we are using effective interactions, which should be
considered as an approximation to the resummed interactions, we can
not determine the actual spin of the parton degrees-of-freedom. For
this reason we neglect the spin of the gluons and treat the quark
spin statistically. In order to describe an equilibration of the
spin degree-of-freedom we introduce a spin flip in 1/3 of the
elastic collisions, as motivated by nuclear physics, that favors
final spin states parallel to the \textbf{B}-field. In order to
simplify the (time expensive) calculations we introduce the
constraint (in equilibrium)
\begin{equation}\label{eq:spinequil}
 n_{\uparrow} P_{\uparrow,\downarrow} = n_{\downarrow} P_{\downarrow,\uparrow} \quad \Longleftrightarrow \quad
\frac{P_{\uparrow,\downarrow}}{P_{\downarrow,\uparrow}}
\end{equation} $$=\frac{n_{\downarrow}}{n_{\uparrow}}=\exp\left(-(E_{\downarrow}-E_
{ \uparrow})/T \right) =\exp \left(-\frac{q e B}{E T} \right),
$$
which is introduced explicitly in the transition matrix element squared.
Accordingly, in (\ref{eq:spinequil}) $P$ denotes the probability for a spin
flip and $n$ the occupation probability for given spin orientation.
In practice the probabilities $P$ are taken as
\begin{equation}\label{eq:spinflipbed}
 P_{\downarrow,\uparrow}=1, \qquad \qquad P_{\uparrow,\downarrow}=\exp \left(-\frac{q e B}{E T} \right)
\end{equation} and lead to the proper equilibrium distribution when
neglecting the $q+\bar{q} \leftrightarrow g$ channels. When
including these channels we find numerically deviations from the
equilibrium distribution by up to 10\% since the gluon channels
reduce the spin orientation in the direction of the ${\bf B}$-field,
i.e. induce a 'diamagnetic effect'.
The actual values of the probabilities (29) are not important for
the present study. As long as they fulfill Eq. (28) the equilibrium
distribution changes only slightly within statistical error bars.
The same holds for the number of collisions with spin flips. Their
values are only important for the timescales of the spin
equilibration, which, however, we do not address.

The magnetization $M$ is defined by the spin density of the system
as
\begin{equation}\label{eq:magnetisierung}
 M=\frac{< \mu_S >}{V} \approx \chi_S e^2 B ,
\end{equation} which in case of small magnetic fields $eB$ - as in our present study - is
proportional to the strength of the $B$-field thus defining a
magnetic susceptibility $\chi_S$ by
\begin{equation}\label{eq12}
 \chi_S=\frac{< \mu_S >}{e^2 B V} .
 \end{equation}

\subsection{Numerical results}
\begin{figure}[htb]
  \centering
  \includegraphics[width=0.95\linewidth]{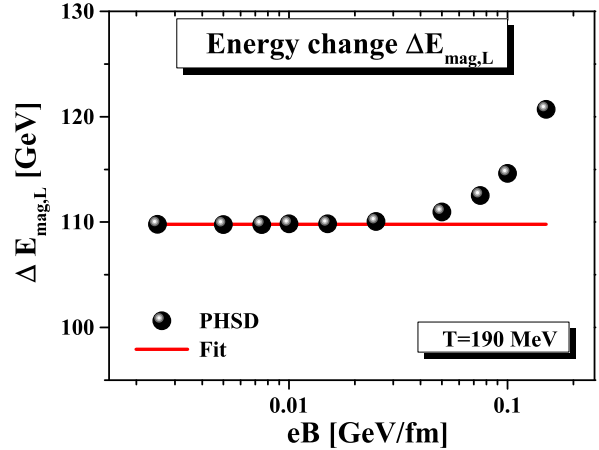}
  \caption{(Color online) The energy shift due to the magnetic field
  $\Delta E_{mag,L}=-\mu_L B$ in the PHSD calculation
  (full dots) as a function of the field strength $eB$ for a temperature
  of $T$ = 190 MeV at $\mu_q$=0. The solid line reflects a constant for small/moderate field strength. }
  \label{pic:mulvsB}
\end{figure}

In order to explore the range of external magnetic fields $eB$ we
can handle reliably within the PHSD calculations for partonic
systems we show in Fig. \ref{pic:mulvsB} the energy contribution to
the magnetic field (\ref{e9}) as a function of $eB$ for a
temperature $T$=190 MeV. In fact, the calculations for the energy
shift due to the magnetic field $eB$ give  constant results for $eB
< $ 50 MeV/fm - when discarding the spin degrees-of-freedom - while
for stronger fields more significant deviations emerge up to $\sim $
10\% for $e B$ 200 MeV/fm. Accordingly, we will restrict to $eB \leq
$ 50 MeV/fm ($\approx$ 0.01 GeV$^2$) in the following.

\begin{figure}[h!tb]
  \centering
  \includegraphics[width=0.95\linewidth]{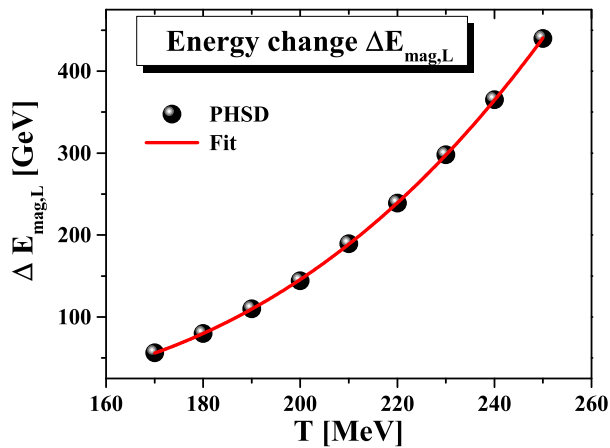}
  \caption{(Color online) The energy shift due to be magnetic field $\Delta E_{mag,L}=-\mu_L B$ in the PHSD calculation
  as a function of the temperature $T$ for $\mu_q=0$. The numerical uncertainties
  are smaller than the size of the dots.
  The solid line shows the fit (\ref{fit5}).}
  \label{pic:muLvsT}
\end{figure}

The temperature dependence of $\Delta E_{mag, L}$ from the PHSD
calculation is shown in Fig. \ref{pic:muLvsT} by full dots and can
be well fitted in the interval 170 MeV $\leq T \leq$ 250 MeV  by
\begin{equation} \label{fit5}
\Delta E_{mag,L}(T)=0.3 \cdot (T-96)^{2.82} \ [MeV]
\end{equation}
where the temperature $T$ is given in units of MeV. The diamagnetic
contribution to the magnetization from the Lorentz force on the
quarks and antiquarks then can be readily extracted by dividing
$\Delta E_{mag, L}(T)$ by the strength of the magnetic field.

\begin{figure}[htb]
  \centering
  \includegraphics[width=0.95\linewidth]{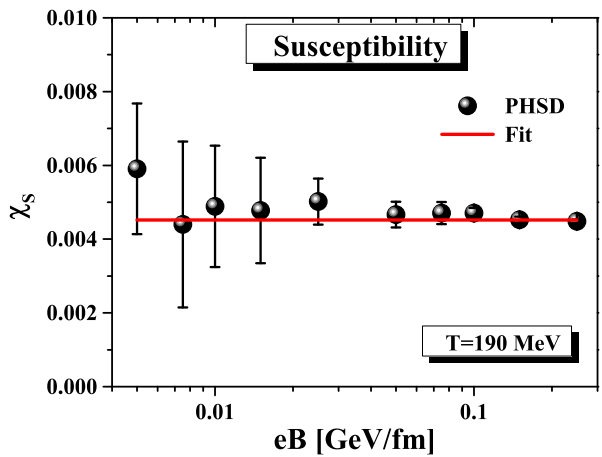}
  \caption{(Color online) Magnetic susceptibility $\chi_S$ (\ref{eq12}) from PHSD as a function of
  the external magnetic field $eB$
for a temperature $T$=190 MeV at vanishing quark chemical potential
$\mu_q$ =0.}
  \label{pic:susceptvsB}
\end{figure}

As a next step we compute the magnetic susceptibility $\chi_S$ in
the PHSD calculations according to Eq. (\ref{eq12}) for different
field strength $eB$ at $\mu_q$ = 0.
We have found the necessary energy for a spin flip to be very small in
comparison to the total energy of the quarks ($< 1\%$) and therefore
have discarded it in our actual simulations.
The results for the susceptibility $\chi_S$ are displayed in Fig.
\ref{pic:susceptvsB} for $T$= 190 MeV and (within numerical
accuracy) show a constant value even up to $eB$= 200 MeV/fm. In this
case the numerical accuracy increases with the field strength since
the spin-flip probabilities in Eq. (\ref{eq:spinflipbed}) differ more
significantly for larger magnetic fields. Nevertheless, we have a
stable numerical 'window' $eB$ from 25-50 MeV/fm where the
diamagnetic and parametic contributions to the magnetic moment can
be calculated with sufficient accuracy. Note that the energy shift
due to the paramagnetic contribution is given by
\begin{equation}
 \Delta E_{mag,S}=-\chi_S V (eB)^2
\end{equation} and decreases quadratically with the field strength.

\begin{figure}[h!tb]
  \centering
  \includegraphics[width=0.95\linewidth]{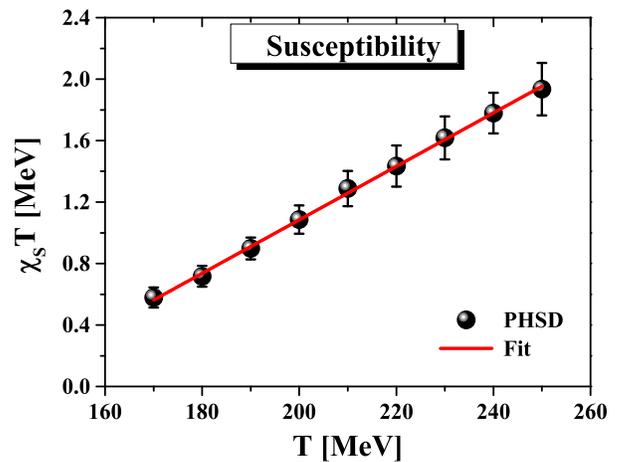}
  \caption{(Color online) Temperature dependence of the magnetic susceptibility
$\chi_S(T) T$ from PHSD (full dots) in comparison to the fit
(\ref{eq13}) for $\mu_q$=0. The numerical uncertainties are
indicated by the errorbars.}
  \label{pic:suszeptvsT}
\end{figure}

The temperature dependence of the magnetic susceptibility
$\chi_S(T)$ from PHSD is displayed in Fig. \ref{pic:suszeptvsT} by
the full dots and can be fitted as
\begin{equation} \label{eq13}
 \chi_S(T)=0.017-\frac{2.39}{T}
\end{equation}
with $T$ given in MeV (in the interval 170 MeV $\leq T \leq$ 250
MeV).

The total energy shift due to the both interactions with the
magnetic field is given by
\begin{equation}
 \Delta E(T,B)=\Delta E_{mag,L}(T) -\chi_S(T) V (eB)^2
\end{equation}
and decreases with $B^2$ at constant temperature $T$. At a
'critical' field $B_c(T)$ the energy shift changes sign, i.e. for
\begin{equation} \label{BBC}
 B_c(T)=\sqrt{\frac{\Delta E_{mag,L}}{e^2 \chi_S V}}
\end{equation}
the magnetization changes from diamagnetic to paramagnetic with
increasing magnitude of the field ${\bf B}$.

\begin{figure}[h!tb]
  \centering
  \includegraphics[width=0.95\linewidth]{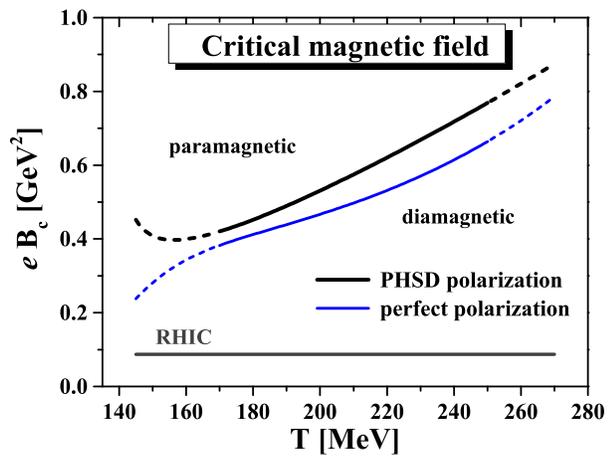}
  \caption{(Color online) The 'critical' magnetic field (\ref{BBC}) as a function of the temperature $T$ for
$\mu_q$
  =0 from the PHSD calculations (thick solid black line). The dashed extensions are based on the extrapolated
  fits and not explicitely controlled by PHSD calculations. The (lower) thin solid blue line results when assuming
  all quark/antiquark spins to be oriented in ${\bf B}$ direction. The constant lower line displays the
  maximum value for the magnetic field strength as found in Ref. \cite{CME7} at the top RHIC energy.
  }
  \label{pic:BcvsT}
\end{figure}

This quantity has a minimum (within PHSD) close to the critical
temperature $T_c \approx $ 158 MeV (cf. Fig. \ref{pic:BcvsT}) with a
minimum $e B_{c,min} \approx 0.4 \ \text{GeV}^2$ (thick solid black line -
extrapolated by the dashed line according to the fits performed).
For comparison we also show the limiting results  when assuming
  all quark/antiquark spins to be oriented in ${\bf B}$ direction.
  This line is slightly lower because the coupling
  to the gluons (in PHSD) reduces the paramagnetic contribution to the magnetization
  (diamagnetic gluon effect).
In the QGP phase the 'critical' field $B_c$  rises with temperature
and separates the diamagnetic (below) from the paramagnetic response
(above)  of the QGP. Note that the maximal field strength in
peripheral Au + Au collisions at the top RHIC energy $\sqrt{s_{NN}}$
= 200 GeV was found to be $\sim 0.09$ GeV$^2$ \cite{CME7} (constant
solid line) - during the passage time of the nuclei - which is
significantly lower than the 'critical' field in Fig.
\ref{pic:BcvsT}. Accordingly, the response of the QGP in actual
heavy-ion experiments should be diamagnetic. However, for the much
higher field strength explored in  lattice QCD calculations
\cite{lqcdb1,lqcdb2,lqcdb3} for temperatures close to $T_c$ the
response should be paramagnetic.

\subsection{Finite quark chemical potential}
\begin{figure}[htb]
  \centering
  \includegraphics[width=0.95\linewidth]{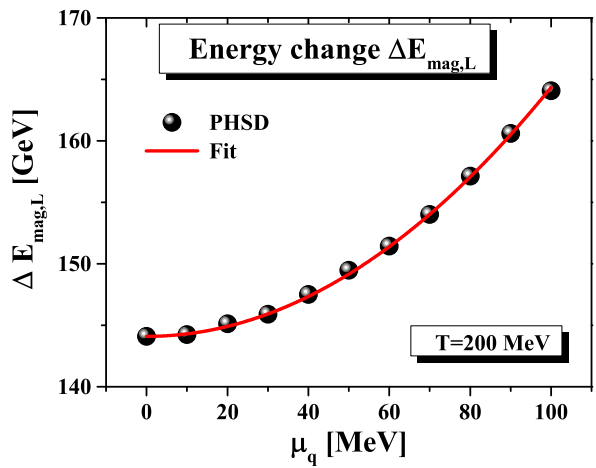}
  \caption{(Color online) The energy shift $\Delta E_{mag,L}(T,\mu_q)$ as a function of $\mu_q$
  from PHSD (full dots) in comparison to the fit (\ref{eq20}) in $\mu_q$ (solid line) for $T$= 200 MeV.}
  \label{pic:muLvsmu}
\end{figure}

As in case of the electric conductivity $\sigma_0(T,\mu_q)$ we can
also compute the magnetic response at finite quark chemical
potential $\mu_q$ in PHSD. In analogy to Fig. 2 we find essentially
a quadratic dependence on $\mu_q$ as demonstrated in Fig.
\ref{pic:muLvsmu} for $\Delta E_{mag,L}(\mu_q)$ at $T$ = 200 MeV.
This dependence is also obtained for the magnetic susceptibility
$\chi_S(T,\mu_q)$ (not shown explicitly) although with larger
numerical errorbars.

\begin{figure}[htb]
   \centering
   \includegraphics[width=0.95\linewidth]{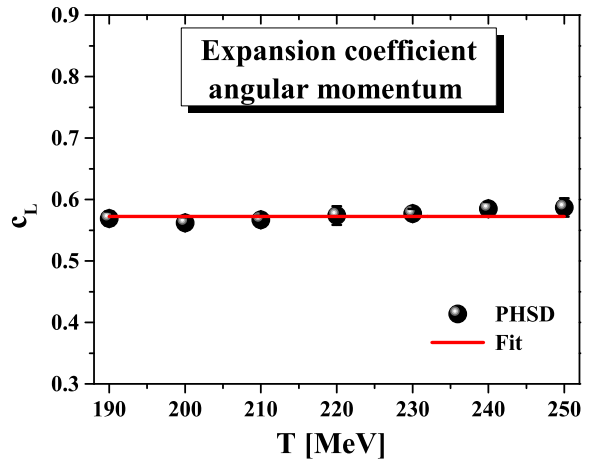}
   \caption{(Color online) The coefficient $c_L$ in Eq. (\ref{eq20}) for $\Delta E_{mag,L}$
   in case of finite $\mu_q \leq$ 100 MeV as a function of temperature $T$.}
   \label{pic:Lkorrfaktor}
\end{figure}

Again we find the temperature dependence of the coefficient to be
$\sim 1/T^{2}$ such that we get the approximations
\begin{equation} \label{eq20}
\Delta E_{mag,L}(T,\mu_q) \approx \Delta E_{mag,L}(T,\mu_q=0) (1+
c_L \frac{\mu_q^2}{T^2}) , \end{equation} $$ \chi_S(T,\mu_q) \approx
\chi_S(T,\mu_q=0) (1+ c_S \frac{\mu_q^2}{T^2}) \ .
$$
As an example we show the coefficient $c_L(T)$ in Fig.
\ref{pic:Lkorrfaktor} for temperatures from 190 to 250 MeV. In this
temperature intervall the expansion coefficient may be well
approximated by $c_L$ = 0.57. Similar statements (with less
accuracy) hold for the magnetic susceptibility in (\ref{eq20}) which
gives $c_S$ = 0.49. The scaling (\ref{eq20}) can be traced back again
to the scaling  of the quark+antiquark density
$n_{q+{\bar q}}(T, \mu_q)$ in the DQPM.

\section{Summary}
In conclusion, we have evaluated the electric conductivity
$\sigma_0(T,\mu_q)$ of the quark-gluon plasma  as a function of
temperature $T$ and quark chemical potential $\mu_q$ by employing
the Parton-Hadron-String Dynamics (PHSD) off-shell transport model
in a finite box for the simulation of dynamical partonic systems in
equilibrium. The PHSD approach in the partonic sector is based on
the lattice QCD equation of state of~\cite{lQCD}; accordingly, it
describes the QGP entropy density $s(T)$, the energy density
$\varepsilon(T)$ and the pressure $p(T)$ from
lQCD~\cite{PHSD1,Bratkovskaya:2011wp,Vitalii1} very well. Studies of
the QCD matter within PHSD have previously given reasonable results
also for the shear and bulk viscosities $\eta$ and $\zeta$ versus
$T$~\cite{Vitalii2} and related transport coefficients \cite{Marty}.

In extension of our previous study in Ref. \cite{Ca13} we have found
that the electric conductivity shows a simple scaling with
$\mu_q^2/T^2$ (cf. Eq.(8)) which essentially can be traced back to
the variation of the quark+antiquark density $n_{q+{\bar q}}$ of the
system in the Dynamical QuasiParticle Model (DQPM). We recall that
PHSD calculations in a fixed box in equilibrium give practically the
same results for the equation of state of QCD matter and transport
coefficients as the DQPM \cite{Vitalii1,Vitalii2}. In analogy to the
ratio of shear viscosity to entropy density $\eta/s$ we find a clear
minimum of the dimensionless ratio $\sigma_0/T$ in the vicinity of
the critical temperature $T_c$ with values close to more recent
lattice QCD calculations \cite{l1,l4,l5}. This prediction as well as
the explicit dependence on $\mu_q$ should be controlled by related
QCD  studies on the lattice. The frequency dependence of the complex
electric conductivity $\sigma(\Omega)$ is found to be well in line
with the simple results from the DQPM (11), too.

Furthermore, we have explored the response of the partonic system on
a moderate external magnetic field $eB \leq 0.1$ GeV/fm and studied
separately the diamagnetic response from the Lorentz force (1) and
the paramagnetic response due to the interaction of the quark spin
with the magnetic field in Eq. (13). Within the temperature range
investigated (170 MeV $< T <$ 250 MeV) the magnetic moment due to
the Lorentz force rises with temperature $T$ but drops with $1/B$
providing a constant energy shift due to a coupling to the magnetic
field $\Delta E_{mag,L}(\mu_q) > 0$. The coupling of the spin to the
$B$-field gives a paramagnetic contribution which can well be
characterized by a magnetic susceptibility $\chi_S(T,\mu_q)$. Its
contribution to the energy shift of the system is negative and
increases with $B^2$. According, there is a 'critical' magnetic
field $B_c(T)$ for which the response of the system changes from
diamagnetic to paramagnetic. The actual values for $B_c(T)$ (cf.
Fig. 12) demonstrate that the response of the QGP in
ultrarelativistic heavy-ion collisions should be diamagnetic since
the maximal magnetic fields created in these collisions are smaller
than $B_c(T)$. As in case of the electric conductivity the
dependence of the energy shift $\Delta E_{mag,L}(\mu_q)$ as well as
the magnetic susceptibility $\chi_S(T,\mu_q)$ show a very similar
scaling correction $\sim \mu_q^2/T^2$ with a coefficient of order
0.5. We close by noting that in the hadronic phase for temperatures
below $T_c$ and $\mu_q$ = 0 the system dominantly consists of
pseudoscalar mesons that have no spin and accordingly a low (or
vanishing) paramagnetic contribution to the magnetization $M$ when
restricting to the first order in the magnetic field. On
the other hand, the charged hadrons see the Lorentz force (1) and
build up a diamagnetic contribution.
\\

The authors acknowledge valuable discussions with E. L. Bratkovskaya, V. Konchakovski,
O. Linnyk and R. Marty during the course of this study which was  supported
by the LOEWE center HIC for FAIR.
\vspace{-0.5cm}


\begin{thebibliography}{99}
%
%
\bibitem{lQCD} Y. Aoki {\it et al.}, Phys. Lett. B \textbf{643}, 46 (2006); S. Borsanyi {\it et al.},
JHEP \textbf{1009}, 073 (2010); JHEP {\bf 1011}, 077 (2010); JHEP
{\bf 1208}, 126 (2012); Phys. Lett. B \textbf{370}, 99 (2014).
%
\bibitem{Peter} P. Petreczky [HotQCD Collaboration], PoS LATTICE
{ \bf 2012}, 069 (2012); AIP Conf. Proc. {\bf 1520}, 103 (2013).
%
\bibitem{StrCoupled1}
M.~Gyulassy and L.~D.~McLerran, Nucl. Phys. A {\bf 750}, 30 (2005);
%
E.~V.~Shuryak, Nucl. Phys. A {\bf 750}, 64 (2005);
%
U.~W.~Heinz, AIP Conf. Proc. {\bf 739}, 163 (2004);
%
A.~Peshier and W.~Cassing, Phys. Rev. Lett. {\bf 94}, 172301 (2005).
%
\bibitem{STAR}
J.~Adams {\it et al.} (STAR Collaboration), Nucl. Phys. A {\bf 757},
102 (2005).
%
\bibitem{PHENIX}
K.~Adcox {\it et al.} (PHENIX Collaboration), Nucl. Phys. A {\bf
757}, 184 (2005).
%
\bibitem{BRAHMS}
I.~Arsene {\it et al.} (BRAHMS Collaboration), Nucl. Phys. A {\bf
757}, 1 (2005).
%
\bibitem{PHOBOS}
B.~B.~Back {\it et al.} (PHOBOS Collaboration), Nucl. Phys. A {\bf
757}, 28 (2005).
%
\bibitem{ALICE}
K.~Aamodt {\it et al.} (ALICE Collaboration), Phys. Rev. Lett. {\bf
105}, 252302 (2010).
%
\bibitem{IdealHydro1}
P.~Huovinen {\it et al.},
Phys. Lett. B {\bf 503}, 58 (2001).
%
\bibitem{IdealHydro2}
P.~F.~Kolb, P.~Huovinen, U.~Heinz, and H.~Heiselberg, Phys. Lett. B
{\bf 500}, 232 (2001).
%
\bibitem{IdealHydro3}
D.~Teaney, J.~Lauret, and E.~V.~Shuryak, Phys. Rev. Lett. {\bf 86},
4783 (2001).
%
\bibitem{IdealHydro4}
T.~Hirano and K.~Tsuda, Phys. Rev. C {\bf 66}, 054905 (2002).
%
\bibitem{IdealHydro5}
P.~F.~Kolb and R.~Rapp, Phys. Rev. C {\bf 67}, 044903 (2003).
%
\bibitem{IdealHydro6}
P.~Huovinen, in {\it Quark-Gluon Plasma 3}, edited by R.~C.~Hwa and
X.-N.~Wang (World Scientific, Singapore, 2004); P.~F.~Kolb and
U.~W.~Heinz, edited by R.~C.~Hwa and X.-N.~Wang (World Scientific,
Singapore, 2004).
%
\bibitem{l0} H. B. Meyer, Phys. Rev. D {\bf 76}, 101701 (2007).
%
\bibitem{ll0} S. Sakai and A. Nakamura, Pos {\bf LAT2007}, 221
(2007).
%
\bibitem{Vitalii1}
V.~Ozvenchuk, O.~Linnyk, M.~I.~Gorenstein, E.~L.~Bratkovskaya, and
W.~Cassing,
Phys. Rev. C {\bf 87}, 024901 (2013).
%
\bibitem{Vitalii2}
V.~Ozvenchuk, O.~Linnyk, M.~I.~Gorenstein, E.~L.~Bratkovskaya, and
W.~Cassing, Phys. Rev. C {\bf 87}, 064903 (2013).
%
\bibitem{review} R. A. Lacey and A. Taranenko, PoS {\bf CFRNC2006}, 021 (2006).
%
\bibitem{MaxBulk1}
D.~Kharzeev and K.~Tuchin, JHEP \textbf{09}, 093 (2008).
%
\bibitem{MaxBulk2}
F.~Karsch, D.~Kharzeev, and K.~Tuchin, Phys. Lett. B {\bf 663}, 217
(2008).
%
\bibitem{MaxBulk3}
P.~Romatschke and D.~T.~Son, Phys. Rev. D {\bf 80}, 065021 (2009).
%
\bibitem{MaxBulk4}
G.~D.~Moore and O.~Saremi, JHEP {\bf 09}, 015 (2008).
%
\bibitem{MaxBulk5}
C.~Sasaki and K.~Redlich, Phys. Rev. C {\bf 79}, 055207 (2009);
Nucl. Phys. A {\bf 832}, 62 (2010).
%
\bibitem{KSS}
G.~Policastro, D.~T.~Son, A.~O.~Starinets, Phys. Rev. Lett. {\bf
87}, 081601 (2001); P.~K.~Kovtun, D.~T.~Son, A.~O.~Starinets, Phys.
Rev. Lett. {\bf 94}, 111601 (2005).

\bibitem{new1} L.P. Csernai, J.I. Kapusta and L.D. McLerran,
Phys. Rev. Lett. 97, 152303 (2006).

\bibitem{new2}
T. Hirano and M. Gyulassy,  Nucl.
Phys. A 769, 71 (2006).
\bibitem{Barbara} B. Jacak and P. Steinberg, Phys. Today {\bf 53},
39 (2010).
%
\bibitem{ViscousHydro1}
P.~Romatschke, U.~Romatschke, Phys. Rev. Lett. {\bf 99}, 172301
(2007).
%
\bibitem{ViscousHydro2}
H.~Song and U.~W.~Heinz, Phys. Rev. C {\bf 77}, 064901 (2008).
%
\bibitem{ViscousHydro3}
M.~Luzum and P.~Romatschke, Phys. Rev. C {\bf 78}, 034915 (2008).
%
\bibitem{ViscousHydro4}
B.~Schenke, S.~Jeon, and C.~Gale, Phys. Rev. C {\bf 82}, 014903
(2010).
%
\bibitem{Greco} S. Plumari, A. Puglisi, F. Scardina, and V. Greco, Phys. Rev. C {\bf 86}, 054902 (2012).
%
\bibitem{Mattiello}
S.~Mattiello and W.~Cassing, Eur. Phys. J. C \textbf{70}, 243
(2010).
%
\bibitem{Hirono:2012rt}
  Y.~Hirono, M.~Hongo and T.~Hirano,
  arXiv:1211.1114.
%
\bibitem{Jorge} S. I. Finazzo and J. Noronha, arXiv:1311.6675.
%
\bibitem{l1} H.-T. Ding {\it et al.}, Phys. Rev. D {\bf 83}, 034504
(2011); O. Kaczmarek {\it et al.}, PoS Confinement X, 185 (2012).

\bibitem{l2} G. Aarts, C. Allton, J. Foley, S. Hands, and S. Kim,
Phys. Rev. Lett. {\bf 99}, 022002 (2007).
\bibitem{l3} S. Gupta, Phys. Lett. B {\bf 597}, 57 (2004).
\bibitem{l4} P. V. Buividovich {\it et al.}, Phys. Rev. Lett. {\bf
105}, 132001 (2010).
\bibitem{l5} 
  B.~B.~Brandt, A.~Francis, H.~B.~Meyer and H.~Wittig,
  PoS Confinement X, 112 (2012). 
%

\bibitem{Tuchin} K. Tuchin,  Adv. High Energy Phys. {\bf 2013}, 490495 (2013)
%
\bibitem{CME1}  D. E.Kharzeev, L. D. McLerran, and H. J.Warringa, Nucl. Phys.
A {\bf 803}, 227 (2008).
\bibitem{CME2}
D. E. Kharzeev, Ann. Phys. (NY) {\bf 325}, 205 (2010).
\bibitem{CME3}
K. Fukushima, D. E. Kharzeev, and H. J. Warringa, Phys. Rev. D {\bf
78}, 074033 (2008).
%
\bibitem{CME4}
D. Kharzeev and A. Zhitnitsky, Nucl. Phys. A {\bf 797}, 67 (2007).
\bibitem{CME5}
D. E. Kharzeev and H. J. Warringa, Phys. Rev. D {\bf 80}, 034028
(2009).
\bibitem{CME6}
V. Skokov, A. Illarionov, and V. Toneev, Int. J. Mod. Phys. A {\bf
24}, 5925 (2009).
%
\bibitem{lqcdb0}
G.S. Bali, F. Bruckmann, G. Endrodi, Z. Fodor, S.D. Katz, S. Krieg, A. Sch\"afer, and K.K. Szabo,
 JHEP {\bf 1202}, 044 (2012).
\bibitem{lqcdb1} G. S. Bali, F. Bruckmann, M. Constantinou, M. Costa, G. Endrodi, S. D. Katz, H. Panagopoulos and A.
Sch\"afer, Phys. Rev. D {\bf 86}, 094512 (2012).

\bibitem{lqcdb2}
G.S. Bali, F. Bruckmann, G. Endrodi, F. Gruber, and A. Sch\"afer,
 JHEP {\bf 1304}, 130 (2013).


\bibitem{lqcdb3}
G.S. Bali, F. Bruckmann, G. Endrodi, and A. Sch\"afer,
e-Print: arXiv:1310.8145

\bibitem{CME7}
V. Voronyuk , V. D. Toneev, W. Cassing, E. L. Bratkovskaya, V.
P.Konchakovski, and S. A.Voloshin, Phys. Rev. C {\bf 83}, 054911
(2011).

\bibitem{Slava1}
V.D. Toneev, V.P. Konchakovski, V. Voronyuk, E.L. Bratkovskaya, and W. Cassing,
 Phys. Rev. C {\bf 86}, 064907 (2012).

\bibitem{photon1}
H. van Hees, C. Gale, and R. Rapp,
Phys. Rev. C {\bf 84}, 054906 (2011).

\bibitem{photon2}
C. Shen, U. Heinz, J.-F. Paquet, and C. Gale, e-Print:
arXiv:1308.2440.

\bibitem{photon3}
O. Linnyk, V.P. Konchakovski, W. Cassing, E.L. Bratkovskaya,
Phys. Rev. C {\bf 88}, 034904 (2013).

\bibitem{Ca13} W. Cassing, O. Linnyk, T. Steinert, and V.
Ozvenchuk, Phys. Rev. Lett. {\bf 110}, 182301 (2013).

\bibitem{Marty}
R. Marty, E. Bratkovskaya, W. Cassing, J. Aichelin, and H. Berrehrah,
Phys. Rev. C {\bf 88}, 045204
(2013).


\bibitem{PHSD1}
W.~Cassing and E.~L.~Bratkovskaya, Nucl. Phys. A \textbf{831}, 215
(2009); Phys. Rev. C \textbf{78}, 034919
(2008).
%
\bibitem{Kadanoff1}
L.~P.~Kadanoff and G.~Baym, {\it Quantum Statistical Mechanics},
(Benjamin, New York, 1962).
%
\bibitem{Kadanoff2}
S.~Juchem, W.~Cassing, and C.~Greiner, Phys. Rev. D \textbf{69},
025006 (2004); Nucl. Phys. A \textbf{743}, 92 (2004).
%
\bibitem{CBRep98}
W. Cassing and E. L. Bratkovskaya, Phys. Rept. {\bf 308}, 65
(1999).
%
\bibitem{Brat97}
E.~L.~Bratkovskaya and W.~Cassing, Nucl. Phys. A {\bf 619}, 413
(1997).
%
\bibitem{Cass02} W. Cassing, Nucl. Phys. A {\bf 700}, 618 (2002).


\bibitem{PRL03}  E. L. Bratkovskaya, S. Soff, H. St\"ocker, M. van Leeuwen,
and W. Cassing,
Phys. Rev. Lett. {\bf 92}, 032302 (2004).

\bibitem{Cass90} W. Cassing,  V. Metag, U. Mosel, and K. Niita,
Phys. Rep. \textbf{188}, 363 (1990).
%
\bibitem{DQPM1}
W.~Cassing, Nucl. Phys. A \textbf{795}, 70 (2007).
%
\bibitem{DQPM2} W.~Cassing, Nucl. Phys. A \textbf{791}, 365
(2007).
%
\bibitem{DQPM3}
A.~Peshier, Phys. Rev. D \textbf{70}, 034016 (2004); J. Phys. G
\textbf{31}, S371 (2005).
%
\bibitem{Bratkovskaya:2011wp}
E.~L.~Bratkovskaya, W.~Cassing, V.~P.~Konchakovski, and O.~Linnyk,
Nucl. Phys. A \textbf{856}, 162 (2011).
%
\bibitem{Cassing}
W.~Cassing, Eur. Phys. J. ST {\bf 168}, 3 (2009).

%
\bibitem{Aarts}
A. Amato, G. Aarts, C. Allton, P. Giudice, S. Hands, and J.-I. Skullerud,
 Phys. Rev. Lett. {\bf 111}, 172001 (2013).
%
\bibitem{Olena13}  O. Linnyk,  W. Cassing, J. Manninen, E.L. Bratkovskaya, P.B. Gossiaux,
J. Aichelin, T. Song, C.M. Ko, Phys. Rev. C {\bf 87}, 014905 (2013).

\end{thebibliography}
\end{document}